%% file: WALpaper.tex
\documentclass[aps,prl,reprint,superscriptaddress, showkeys]{revtex4-2}

\input{declarations}

\begin{document}

\input{newcommands}

\title{Boosting proximity spin orbit coupling in graphene/WSe$_2$ heterostructures via hydrostatic pressure}
\input{authors}


\input{sec_abstract} 

\date{\today}
\maketitle

\input{sec_intro}

\input{sec_meas}

\input{sec_pdep}

\input{sec_curvefit}

\input{sec_conclusion}

\input{sec_ack_contrib}

\input{references}
 
\end{document}

\input{newcommands}

\beginsupplement

\title{Supplemental Material: Boosting proximity spin orbit coupling in graphene/WSe$_2$ heterostructures via hydrostatic pressure}

\input{authors}

\date{\today}
\maketitle

\input{SM_raw}

\include{SM_compare}

\include{SM_tdep}

\include{SM_fitting}

\input{SM_otherdevice}

\input{references}

%% file: declarations.tex
\usepackage[utf8]{inputenc}
\usepackage[T1]{fontenc}
\usepackage{amsmath, amssymb, graphicx, lmodern}

%% file: newcommands.tex
% straight unit letters with small space in math mode. hyphens are minus signs.
\newcommand{\unit}[1]{\,\mathrm{#1}}
% straight font for upright indexing and text, hyphens are hyphens, etc.
\newcommand{\tn}[1]{\textnormal{#1}}

% In order to label all fig. with S1, S2,... in the SuppMat
\newcommand{\beginsupplement}{%
        \setcounter{table}{0}
        \renewcommand{\thetable}{S\arabic{table}}%
        \setcounter{figure}{0}
        \renewcommand{\thefigure}{S\arabic{figure}}%
     }

% Parameter names
\newcommand{\LIA}{\lambda_\tn{I}^\tn{A}}
\newcommand{\LIB}{\lambda_\tn{I}^\tn{B}}
\newcommand{\LR}{\lambda_\tn{R}}
\newcommand{\LVZ}{\lambda_\tn{VZ}}
\newcommand{\LPIAA}{\lambda_\tn{PIA}^\tn{A}}
\newcommand{\LPIAB}{\lambda_\tn{PIA}^\tn{B}}
\newcommand{\LPIA}{\lambda_\tn{PIA}}

\newcommand{\tbi}{\tau_B^{-1}}
\newcommand{\tphi}{\tau_\varphi^{-1}}
\newcommand{\tivi}{\tau_\tn{iv}^{-1}}
\newcommand{\tiai}{\tau_\tn{ia}^{-1}}
\newcommand{\tai}{\tau_\tn{asy}^{-1}}
\newcommand{\tsi}{\tau_\tn{sym}^{-1}}
\newcommand{\tsti}{\tau_\tn{*}^{-1}}
\newcommand{\twi}{\tau_\tn{w}^{-1}}
\newcommand{\tmi}{\tau_\tn{m}^{-1}}

\newcommand{\tb}{\tau_B}
\newcommand{\tph}{\tau_\varphi}
\newcommand{\tiv}{\tau_\tn{iv}}
\newcommand{\tia}{\tau_\tn{ia}}
\newcommand{\ta}{\tau_\tn{asy}}
\newcommand{\ts}{\tau_\tn{sym}}
\newcommand{\tst}{\tau_\tn{*}}
\newcommand{\tw}{\tau_\tn{w}}
\newcommand{\tm}{\tau_\tn{m}}

\newcommand{\bph}{B_\varphi}
\newcommand{\biv}{B_\tn{iv}}
\newcommand{\bia}{B_\tn{ia}}
\newcommand{\ba}{B_\tn{asy}}
\newcommand{\bs}{B_\tn{sym}}
\newcommand{\bst}{B_\tn{*}}
\newcommand{\bw}{B_\tn{w}}
\newcommand{\bmom}{B_\tn{m}}

%% file: authors.tex
\author{Bálint Fülöp}
\affiliation{Department of Physics, Budapest University of Technology and Economics and Nanoelectronics ``Momentum'' Research Group of the Hungarian Academy of Sciences, Budafoki út 8, 1111 Budapest, Hungary}

\author{Albin Márffy}
\affiliation{Department of Physics, Budapest University of Technology and Economics and Nanoelectronics ``Momentum'' Research Group of the Hungarian Academy of Sciences, Budafoki út 8, 1111 Budapest, Hungary}

\author{Simon Zihlmann}
\affiliation{Department of Physics, University of Basel, Klingelbergstrasse 82, CH-4056 Basel, Switzerland}

\author{Martin Gmitra}
\affiliation{Institute of Physics, Pavol Jozef Šafárik University in Košice, Park Angelinum 9, 040 01 Košice, Slovak Republic}

\author{Endre Tóvári}
\affiliation{Department of Physics, Budapest University of Technology and Economics and Nanoelectronics ``Momentum'' Research Group of the Hungarian Academy of Sciences, Budafoki út 8, 1111 Budapest, Hungary}

\author{Bálint Szentpéteri}
\affiliation{Department of Physics, Budapest University of Technology and Economics and Nanoelectronics ``Momentum'' Research Group of the Hungarian Academy of Sciences, Budafoki út 8, 1111 Budapest, Hungary}

\author{Máté Kedves}
\affiliation{Department of Physics, Budapest University of Technology and Economics and Nanoelectronics ``Momentum'' Research Group of the Hungarian Academy of Sciences, Budafoki út 8, 1111 Budapest, Hungary}

\author{Kenji Watanabe}
\affiliation{Research Center for Functional Materials, National Institute for Materials Science, 1-1 Namiki, Tsukuba 305-0044, Japan}

\author{Takashi Taniguchi}
\affiliation{International Center for Materials Nanoarchitectonics, National Institute for Materials Science,  1-1 Namiki, Tsukuba 305-0044, Japan}

\author{Jaroslav Fabian}
\affiliation{Institute for Theoretical Physics, University of Regensburg, 93040 Regensburg, Germany}

\author{Christian Sch\"onenberger}
\affiliation{Department of Physics, University of Basel, Klingelbergstrasse 82, CH-4056 Basel, Switzerland}

\author{P\'eter Makk}
\email{peter.makk@mail.bme.hu}
\affiliation{Department of Physics, Budapest University of Technology and Economics and Nanoelectronics ``Momentum'' Research Group of the Hungarian Academy of Sciences, Budafoki út 8, 1111 Budapest, Hungary}

\author{Szabolcs Csonka}
\email{csonka@mono.eik.bme.hu}
\affiliation{Department of Physics, Budapest University of Technology and Economics and Nanoelectronics ``Momentum'' Research Group of the Hungarian Academy of Sciences, Budafoki út 8, 1111 Budapest, Hungary}

%% file: sec_abstract.tex
\begin{abstract}
Van der Waals heterostructures composed of multiple few layer crystals allow the engineering of novel materials with predefined properties.
As an example, coupling graphene weakly to materials with large spin orbit coupling (SOC) allows to engineer a sizeable SOC in graphene via proximity effects.
The strength of the proximity effect depends on the overlap of the atomic orbitals, therefore, changing the interlayer distance via hydrostatic pressure can be utilized to enhance the interlayer coupling between the layers.
In this work, we report measurements on a graphene/WSe$_2$ heterostructure exposed to increasing hydrostatic pressure.
A clear transition from weak localization to weak anti-localization is visible as the pressure increases, demonstrating the increase of induced SOC in graphene.
\end{abstract}

%% file: sec_intro.tex
Graphene based van der Waals (vdW) heterostructures became one of the most studied physical systems in material science in recent years, which led to the emergence of designer electronics \cite{Geim2013, Giustino2021}.
Since the electrons are localized at the surface for a single layer of graphene by definition, their properties can be easily modified by combining it with other few layer crystals leading to remarkable changes in its band structure. 
A prominent example is the moiré effect caused by the rotation (and possible small lattice mismatch) of the graphene and the underlying other lattice. 
This led to the Hofstadter physics and formation of secondary charge neutrality points (CNPs) when graphene is placed on hexagonal boron nitride (hBN) \cite{Dean2013, Ponomarenko2013, Hunt2013, KrishnaKumar2017, KrishnaKumar2018, Wang2019, Wang2019a, Yankowitz2019a}; whereas correlated phases including superconductivity, correlated insulators or ferromagnetic states have been found if it is placed on another graphene sheet \cite{Cao2018a, Cao2018b, Sharpe2019, Lu2019}. 
Graphene based heterostructures are also  promising building blocks for spintronic devices \cite{Han2014, Avsar2020, Hu2020}.

\begin{figure}
\includegraphics[width=9cm]{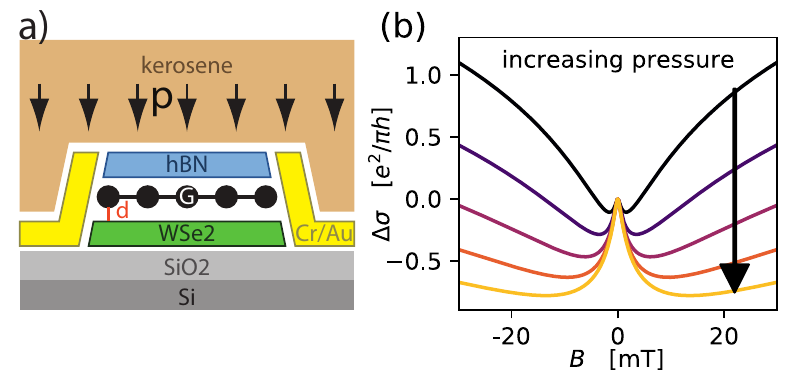}
\caption{
(a)
Schematic side view of a hBN/graphene/WSe$_2$ heterostructure in kerosene pressure transfer medium.
Hydrostatic pressure reduces the distance $d$ between the graphene and the WSe$_2$ layers (among others), which leads to an enhancement of the proximity-induced SOC in graphene.
(b)
Simulated weak anti-localization curves using realistic parameters to demonstrate the potential effect of the application of ca. 2\,GPa pressure on the heterostructure.
The increased SOC leads to a more pronounced WAL peak in the magneto-conductivity curve.
See the Supp.\ Mat.\ for the simulation details.
}
\label{fig:intro}
\end{figure}

Although graphene is known to provide very long spin lifetimes \cite{Kamalakar2015, Droegeler2016}, the absence of spin orbit coupling (SOC) also hinders electrical control and charge to spin conversion in it.
However, a large SOC can be induced in graphene by proximity effect if placed on a transition metal dichalcogenide (TMDC) flake \cite{Gmitra2015, Gmitra2016, Garcia2017}, which can lead to topologically nontrivial states and the quantum spin Hall effect \cite{Kane2005}. 
Recently, a wide range of experiments demonstrated the presence of proximity-induced SOC in various heterostructures by weak-localization, capacitance or spin transport measurements, and it has been found that Rashba and valley--Zemann-like SOC is induced in graphene leading to a large spin relaxation anisotropy \cite{Avsar2014, Wang2015, Wang2016, Yang2016, Ghiasi2017, Yang2017, Benitez2018, Zihlmann2018, Ringer2018, Island2019, Wakamura2019, Amann2012.05718}.
Since this enhancement of SOC originates from the hybridization of graphene's $\pi$ orbitals with the TMDC layer's outer orbitals, the strength of the SOC depends strongly on the overlap of the orbital wavefunctions and, therefore, on the interlayer distance, which is determined by the van der Waals force.
Compressing such a heterostructure by applying an external pressure (see Figure \ref{fig:intro}a), is expected to increase the SOC, which can be captured by weak localization measurements, as shown by simulated magneto-conductivity curves in Figure \ref{fig:intro}b \cite{Gmitra2016, carr2018}.

In this work we present experimental evidence of manipulation of the interlayer coupling and the proximity SOC in a graphene/WSe$_2$ heterostructure using hydrostatic pressure.
The pressure control adds another knob with which the electronic properties of 2D materials can be engineered, allowing to more robust proximity states or even engineer novel states of matter.

%% file: sec_meas.tex
\begin{figure}
\includegraphics[width=8.8cm]{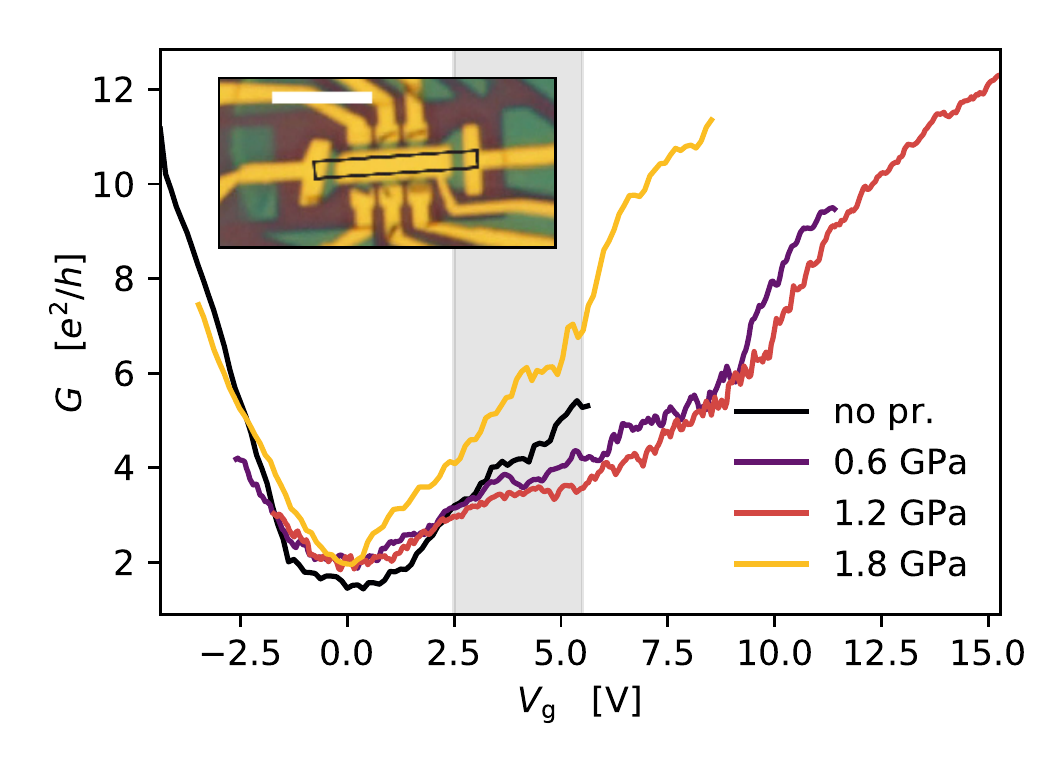}
\caption{
Two-terminal conductance vs. back gate voltage measurements at different pressures. 
The curve minima, assumed to be the CNP, are shifted to $V_\tn{g}=0$ to maintain comparability of the curve shapes.
The area considered for the comparison of the WAL signals, between $2.5\unit{V}$ and $5.5\unit{V}$, is highlighted by the gray background.
Inset:
Optical micrograph of the sample. 
Scale bar is 10\,\textmu{}m. 
The highlighted segment is measured in two-terminal measurements at 1.5\,K for all pressures.
}
\label{fig:sample}
\end{figure}

The illustration of the studied device is shown in Figure \ref{fig:intro}a, whereas an optical image is given in the inset of Figure \ref{fig:sample}.
The heterostructure is built on a Si/SiO$_2$ substrate using the dry stacking assembly method \cite{Zomer2014}, where the heavily-doped silicon layer was used as a global back gate.
It consists of a monolayer graphene, which is on top of a very thin (3\,nm) WSe$_2$ flake providing the spin orbit coupling, and covered with a hexagonal boron nitride (hBN) flake to protect it from the kerosene pressure medium \cite{Fulop2021tech} (see Figure \ref{fig:intro}a).
The heterostructure is shaped into a Hall bar and contacted using 1D Cr/Au electrodes \cite{Wang2013} (see Figure \ref{fig:sample} inset).
The total length of the graphene segment was $L = 8.3$\,\textmu{}m, and the width was $W = 1.1$\,\textmu{}m.
The device is also equipped with a top gate extending over the major part of the measured segment but it was grounded during the measurements.
Another sample, showing similar behaviour is shown in the Supp.\ Mat.

The wafer carrying the device was cut tightly and bonded on a special high pressure sample holder, then placed in kerosene environment in a  piston-cylinder hydrostatic pressure cell.
The setup is  designed to overcome the technical difficulties of electronic
measurements of nanocircuits in a hostile environment \cite{Fulop2021tech}.
Low temperature (1.5\,K) measurements have been carried out at four different hydrostatic pressure settings in increasing order (no pressure, 0.6\,GPa, 1.2\,GPa, 1.8\,GPa).
Each pressure change involved warming up the sample to room temperature, applying the pressure using a hydraulic press, clamping the pressure cell, and cooling the sample down again.
Measurements were carried out by standard lock-in technique at $f=177\unit{Hz}$ with an AC bias voltage $V_\mathrm{AC} = 100$\,\textmu{}V and an external low-noise current amplifier.

We characterized our devices by measuring the two-terminal conductance as a function of back gate voltage at each pressure, as shown in Figure \ref{fig:sample}.
The measured segment is highlighted in the inset.
The CNP position was found at slightly different $V_\tn{g}$ back gate voltages in each case but remained between -5.2\,V and -0.6\,V, which was corrected by shifting the curve minima to $V_\tn{g} = 0$ for further use.
The curves are quite similar in shape, with a minimum around the CNP.
Using a simple parallel plate capacitor model for the estimation of the charge carrier density $n(V_\tn{g})$, field effect mobility was calculated based on a linear fit on the two-terminal conductance.
Electron mobility values were found between $11,000$ and $24,000\unit{cm^2 V^{-1}s^{-1}}$ without any systematic dependence on the applied pressure.
A change in the scattering processes and the observed field effect mobility is not unusual in case of vdW heterostructures during subsequent cooldowns even without applied pressure and can be attributed to the rearrangement of scattering centers.
Thus we conclude that the sample conductance and quality is not significantly affected by the applied pressure.

%% file: sec_pdep.tex
\begin{figure*}
\includegraphics[width=\textwidth]{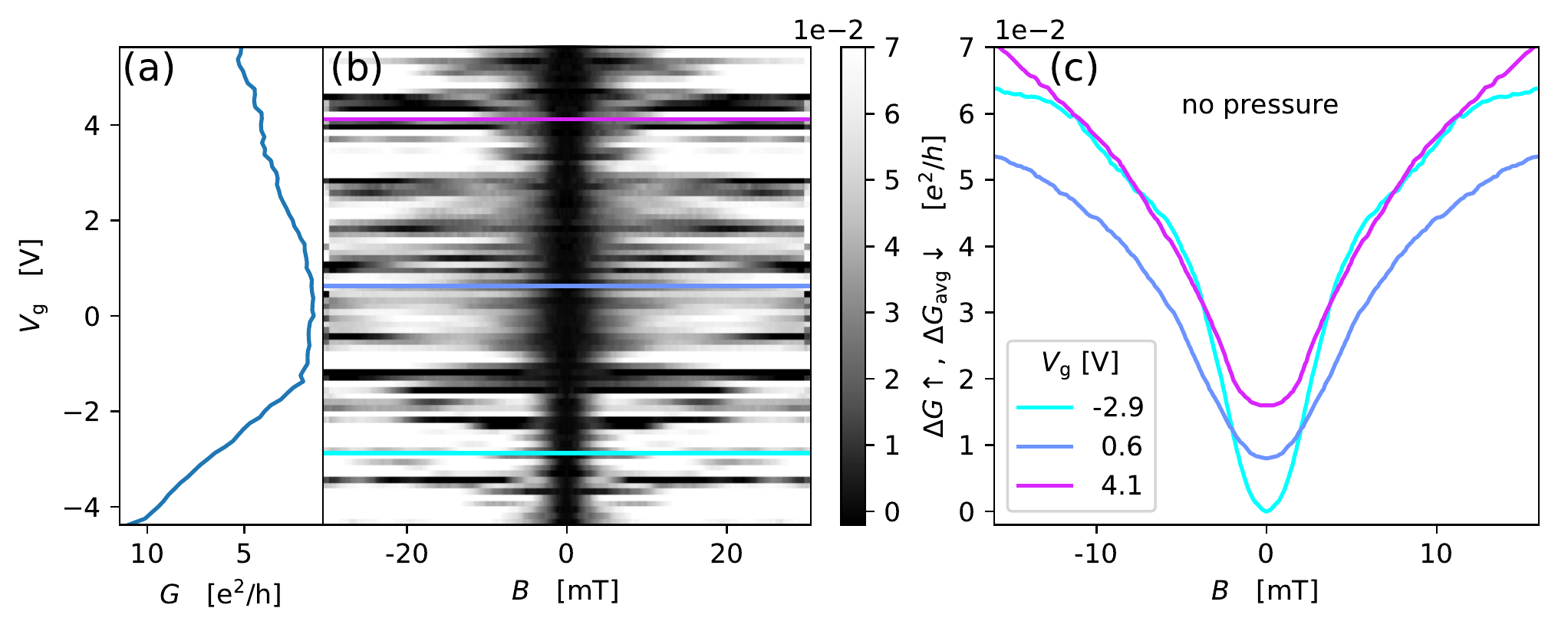}
\caption{
	Weak localization measurement at ambient pressure.
	(a) 
	Zero-field conductance $G(V_\tn{g}, B=0)$, used to extract $D$ and $\tia$, as detailed in the text. 
	(b) 
	2D grayscale plot of the two-terminal conductance corrected by the zero-field conductance $\Delta G(V_\tn{g}, B) = G(V_\tn{g},B) - G(V_\tn{g}, B=0)$. 
	(c)
	Two-terminal magneto-conductance at fixed gate voltages marked in panel\,b with $\pm 1.5 \unit{V}$ averaging along the vertical axis to reduce the effect of UCF.
	The curves are shifted by 0.01\,$e^2$/h for clarity.
	At small magnetic fields $(B < 5 \unit{mT})$ the well-formed conductance dip corresponds to weak localization (WL) effect that is present at all gate voltages, although the curve shape changes slightly.
	At higher fields $(B > 5 \unit{mT})$ the influence of UCF leads to irregular line shapes and were not analysed in the current work.
}
\label{fig:rainbow}
\end{figure*}

Now we turn to low-field magneto-conductance measurements.
Weak localization is a low temperature quantum correction to the magneto-conductance of diffusive samples and expected to show a conductance minimum (weak localization, WL) or maximum (weak anti-localization, WAL) at zero field depending on the strength of SOC\cite{IhnBook2004}.

We have recorded the two-terminal conductance curves as the B field was swept in a range of $\pm 30 \unit{mT}$ for several gate voltages, both in the up and down magnet ramping direction.
At each gate voltage, we symmetrized the curves, then subtracted the zero-field conductance from the measured curve leading to 2D conductance maps $\Delta G(V_\tn{g},B)$.
The zero-field conductance against the gate voltage for the no pressure case is plotted in Figure \ref{fig:rainbow}a.
The corresponding conductance map is shown in Figure \ref{fig:rainbow}b, where a vertical dark gray line along $B = 0$ appears due to the correction method, and lighter gray tones on each side are a sign of positive magneto-conductance corresponding to WL.
At higher magnetic fields, universal conductance fluctuations \cite{IhnBook2004} (UCF) of amplitude up to $0.4\unit{e^2/h}$ across the gate voltage axis are also visible on the map as saturated horizontal lines.
To increase visibility of WAL signal, an averaging on a gate range of $\pm 1.5\unit{V}$ was applied on the conductance, noted as $\Delta G_\tn{avg}(V_\tn{g}, B)$. 
Cuts of $\Delta G_\tn{avg}$ at fixed $V_\tn{g}$ values are shown in fig\,\ref{fig:rainbow}c.
All these cuts show a WL dip with slight changes in the curve shape as a function of the gate voltage.
Here, the absence of a WAL peak suggest a weak SOC, which will be discussed later.
This is in agreement of previous measurements on this device in the Supp.\ Mat.\ of Ref.\,\cite{Zihlmann2018}.

After having performed the above procedure for each pressure, we selected the curves at $4 \pm 1.5 \unit{V}$, i.e.\ averaging the $\Delta G(V_\tn{g}, B)$ curves between $2.5\unit{V}$ and $5.5\unit{V}$ for all the pressures.
We assume the characteristic times of the scattering processes do not change much across the gate range where the averaging is performed, while the contribution of UCF is reduced.
The conductance was converted to conductivity for the extraction of scattering time scales by curve fitting.
Background signal was also recorded simultaneously with the localization measurement and found to be constant, see the Supp.\ Mat.\ for details.

%% file: sec_curvefit.tex
\begin{figure*}
\includegraphics[width=\textwidth]{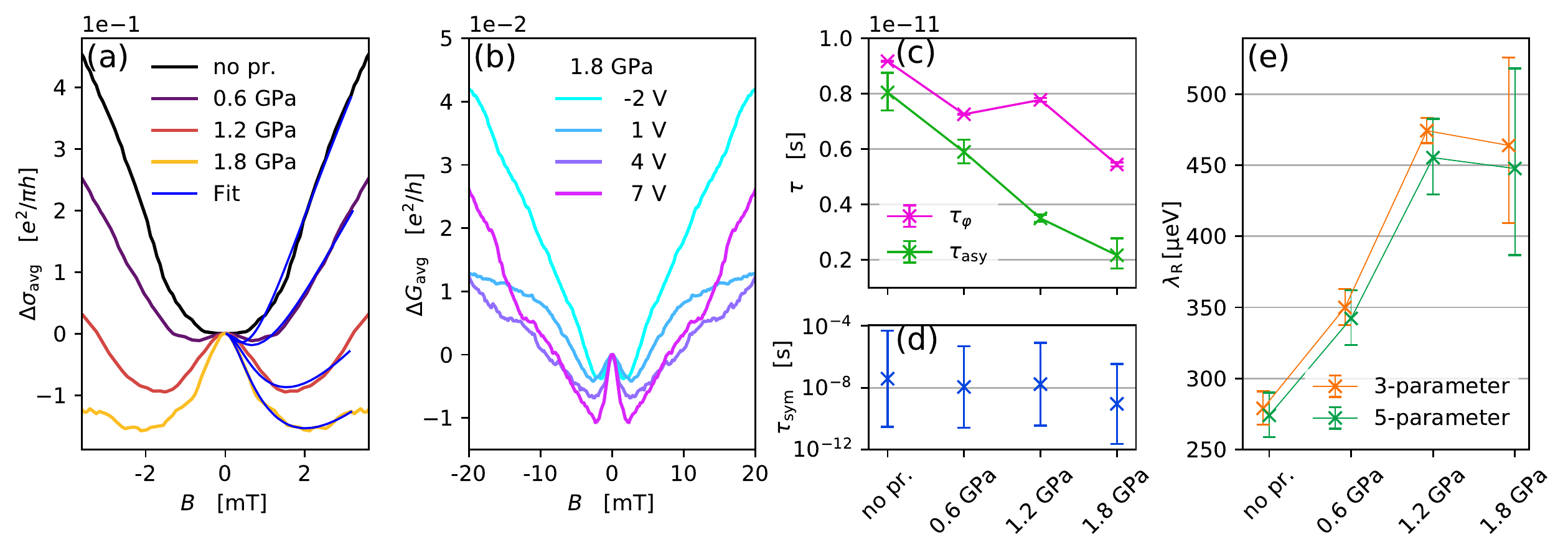}
\caption{
	(a) 
	Comparison of averaged $\Delta \sigma_\tn{avg}(B)$ measurement curves for each pressure at 4\,V.
	Clear signature of the WL $\rightarrow$ WAL evolution is visible. 
	Solid blue lines for $B > 0$: fits using Equation \ref{eq:soc3}.
	(b) 
	Two-terminal magneto-conductance for 1.8\,GPa pressure at fixed gate voltages with $\pm 1.5 \unit{V}$ averaging to reduce the effect of UCF, similarly to \ref{fig:rainbow}c.
	The WAL peak observed in panel\,a is present at all gate values with changing amplitude but the width staying approximately the same.
	At higher fields $(B > 5 \unit{mT})$ the influence of UCF is still dominant.
	(c)
	Summary plot of the fit values of $\tph, \ta$.
	The errorbars represent uncertainty assessed by our method detailed in the Supp.\ Mat.
	Decreasing $\ta$ with increasing pressure indicates the presence of an increasing Rashba SOC and interlayer coupling strength between graphene and WSe$_2$.
	(d)
	Fit values of $\tm$.
	Due to the large uncertainty, the value of this parameter cannot be extracted.
	(e)
	$\LR$ parameter for each pressure, calculated using the previously extracted parameter values for $\ta$ and $\tm$. Calculated values based on the 3-parameter formula (Equation \ref{eq:soc3}) are plotted with orange, and values based on the 5-parameter formula are plotted with green (see the Supp.\ Mat.\ for details).
	A clear growing tendency proves the enhancement of the proximity SOC induced by the WSe$_2$ layer.
	The data points are slightly shifted horizontally to avoid overlapping errorbars.
}
\label{fig:pdep}
\end{figure*}

Our main findings are presented in Figure \ref{fig:pdep}a, where the averaged magneto-conductivity $\Delta\sigma_\tn{avg}(V_\tn{g}=4\unit{V},B)$ for all four pressures is plotted for low magnetic fields.
The initial curve, at ambient pressure, shows a wide conductivity dip, which gets wider as the pressure increases to 0.6\,GPa, and a sharp central peak appears at higher pressures.
This transition of WL to WAL is a clear signature of the increasing proximity-induced SOC in the graphene layer.
This tendency is robust for the entire gate voltage range as shown by the cuts at different gate voltages in Figure \ref{fig:pdep}b.
See the Supp.\ Mat.\ for the full comparison of all pressures at various gate voltages.
To quantitatively demonstrate this transition, the data was fitted using the McCann-Falko weak anti-localization formula \cite{McCann2012}:

\begin{equation}
\begin{split}
\Delta \sigma(B) = -\frac{1}{2} \frac{e^2}{\pi h} 
	\left( 
		F \left( \frac{\tbi}{\tphi} \right)
		- F \left( \frac{\tbi}{\tphi + 2 \tai} \right) \right.\\
		\left.
		- 2 F \left( \frac{\tbi}{\tphi + \tai + \tsi} \right)
	\right),
\end{split}
\label{eq:soc3}
\end{equation}

\noindent{}where $\Delta \sigma(B) = \sigma(B) - \sigma(B=0)$ is the correction to the magneto-conductivity. 
We introduced the function $F (x) = \ln (x) + \psi (0.5 + x^{-1})$ with $\psi$ being the digamma function and $h$ being the Planck's constant. 
The rate $\tbi = 4eDB/\hbar$ is associated to the magnetic field with $D$, the diffusion constant. $\tph$ is the phase-breaking time, $\ta$ is the scattering time due to SOC terms that are asymmetric upon $z$/-$z$ inversion, and $\ts$ is the corresponding time due to terms invariant under $z$/-$z$ inversion.
Using the previously calculated charge carrier density $n(V_\tn{g})$, the Fermi velocity of graphene, and the Einstein relation, the diffusion constant and the momentum relaxation time can be extracted from the gate voltage curves and were found between $D = 0.07$--$0.12\unit{m^2 s^{-1}}$ and $\tm = 1.3$--$2.4\cdot 10^{-13} \unit{s}$ for all pressure measurements without any systematic pressure dependence, but in correlation with the field effect mobility.
Since $\tm$ is expected to be the shortest amongst the time scales, we set it as the lower bound for all fitting time parameters.
Equation \ref{eq:soc3} is valid if the intervalley scattering time $\tiv$ is much shorter than the times corresponding to SOC, therefore the contribution of the effect of the charge carriers' chiral nature can be neglected.
We also performed fits with a more complex, 5-parameter fitting formula including the  $\tiv$ intervalley and $\tia$ intravalley scattering times as fitting parameters, which suggest that $\tiv$ is indeed at least an order of magnitude smaller than the other fitted time scales, supporting the validity of Equation \ref{eq:soc3} (see the Supp.\ Mat.).

In Figure \ref{fig:pdep}a, fit curves are plotted using solid blue lines.
The extracted fit parameters are summarised in Figure \ref{fig:pdep}c-d.
A clear tendency of the reduction of $\ta$ can be observed as its value decreases from $7.4 \cdot 10^{-12}\unit{s}$ to $2.7 \cdot 10^{-12}\unit{s}$ as the pressure increases to 1.8\,GPa.
This is a reduction by a factor of 2.7 and corresponds to an increasing SOC. 
This is the main proof to our initial expectation of increasing interlayer coupling.
The phase-breaking time, $\tph$, also shows a moderately decreasing but fluctuating trend with pressure, obtaining values from $9.2\cdot10^{-12}\unit{s}$ to $5.5\cdot10^{-12}\unit{s}$, staying well above $\ta$ at all pressures.
The reason for this decreasing tendency is unclear at the moment.
The third fit parameter, $\ts$, obtains fit values that are much longer than $\tph$ and the fit errors are more than an order of magnitude large (see Figure \ref{fig:pdep}d), which means that it has negligible effect on the magneto-conductance curve.
Therefore, although there seems to be a decreasing trend in the lifetime, we can not extract any reliable value for it at any pressure.
Discussion on this result will be given later.

The spin relaxation times can be connected to band structure parameters via spin relaxation mechanisms. 
It has been established that two relevant spin orbit terms are formed in graphene/TMDC heterostructures. 
One of them is the Rashba term, described by the Hamiltonian $H_\tn{R}  =  \LR(\kappa\sigma_x s_y-\sigma_y s_x)$ \cite{Konschuh2010}, which corresponds to the breaking of the lateral mirror symmetry due to the difference of the hBN and TMDC neighbouring layers to the graphene sheet.
Its strength is set by the $\LR$ parameter, $\kappa = \pm 1$ for the K and K' valleys, the Pauli matrices $\sigma$ are acting on the lattice pseudospin, and $k_x$, $k_y$ are the electron wave vector components measured from the K (K') points.
This term leads to relaxation processes that are contained in $\ta$ via the Dyakonov--Perel mechanism, from which the Rashba parameter can be expressed as $\LR =  \frac{\hbar }{2} \sqrt{\tmi \tai}$ \cite{Cummings2017}.

The increasing scattering rates lead to an increase in the Rashba parameter, which clearly demonstrates that we were able to tune the strength of the induced SOC. 
The values are summarised in Figure \ref{fig:pdep}e, shown by the orange curve.
We have also plotted the value of the Rashba parameter from the 5-parameter fitting formula in green (see Supp.\ Mat.), which is in agreement with the values from the simplified formula within the error limits.

The second dominant spin orbit term in these systems is the $H_\tn{VZ}  =  \LVZ \kappa \sigma_0 s_z$ valley--Zeeman term, where $\LVZ$ characterizes the coupling strengh.
This term corresponds to an effective Zeeman magnetic field which is opposite in the two valleys due to inversion symmetry breaking, but still preserving time reversal symmetry. 
It leads to relaxation contained in the $\ts$ via a modified Dyakonov--Perel mechanism, where the intervalley scattering time, $\tiv$, enters instead of the momentum scattering time: $\LVZ =  \frac{\hbar }{2} \sqrt{\tivi \tsi}.$

The strength of the valley--Zeeman coupling cannot be extracted reliably in our experiments due to the uncertainty in $\ts$. 
The mean values from the fit show a slightly increasing trend and would give 4\,\textmu{}eV and 23\,\textmu{}eV for the lowest and highest pressure, respectively, but even taking the smallest obtained values from $\ts$, we arrive at values between 140 and 440\,\textmu{}eV.

In order to quantify changes in the Rashba parameter and to compare it to expectations, we have performed ab initio calculations following our previous results in Ref.\,\cite{Gmitra2015, Gmitra2016}.
Using a structural supercell model of $4 \times 4$ graphene and $3 \times 3$ WSe$_2$ with relaxed lateral atomic positions, 5\%/GPa compressibility for the layer distance between the graphene and WSe$_2$ layers was found.
The applied 1.8\,GPa pressure induces 9\% compression in the z direction that leads to an increment of the calculated Rashba energy from 600\,\textmu{}eV to 1.8\,meV and of the calculated valley--Zeeman energy from 1.2\,meV to 3.0\,meV.

As plotted in Figure \ref{fig:pdep}e, the value of the Rashba energy $\LR$ increases from 0.3\,meV to 0.5\,meV, which is in the order of magnitude of the expected values, although slightly lower than them.
The values of the valley--Zeeman energy are much smaller than expected theoretically. 

Several reasons might be behind these low SOC values relative to the theoretical expectations. 
One possibility is that the interfaces are not clean enough and some contamination is trapped in-between, which would lower the strength of the spin orbit coupling. 
Another difference from other samples \cite{Zihlmann2018} might come from the relative orientation of the graphene and the WSe$_2$. 
It has been theoretically found \cite{Li2019, David2019} that the rotation angle strongly modulates the strength of the SOC and for certain angles the SOC almost disappears. 
Since we did not control the twist angles between the layers during fabrication, they may have aligned such a way that it would lead to reduced SOC.
The rotation angle affects the Rashba and the valley--Zeeman coupling in a different way, which can explain why we see stronger suppression for the $\LVZ$ than for $\LR$, and why the former deviated between measured and simulated values. 
Finally, we note that the strength of SOC obtained from these formulas might also depend on how well the UCF is removed during averaging.
This can be seen in Figure \ref{fig:pdep}b, where the different gate values lead to slightly different magneto-conductance trend. 
We stress here that although the absolute value of the extracted parameters have to be taken with care, the tendency visible in Figure \ref{fig:pdep}a clearly shows that the pressure indeed leads to an increase of the SOC.
We do not expect that the twist angle is modulated by the applied pressure \cite{Yankowitz2019}; nor the functional form of the angle dependence is modified by it, only the overall strength of the SOC is affected.

%% file: sec_conclusion.tex
In conclusion, we studied for the first time the effect of hydrostatic pressure on WL signal in graphene/TMDC heterostructure.
We demonstrated the enhancement of the proximity-induced SOC using hydrostatic pressure. 
Our analysis of the measured signals supports an increasing SOC and is in qualitative agreement with theoretical expectations.
The strength of SOC is an important parameter in graphene spintronics, since it determines charge to spin conversion efficiencies and also plays a central role in graphene--TMDC optoelectronic devices \cite{Gmitra2015}.
Our work also points out that hydrostatic pressure can be generally used to change the interlayer distance up to a remarkable 10\%, which provides a new way to boost proximity effects in van der Waals heterstructures, e.g.\ exchange interaction induced into graphene \cite{Ghazaryan2018, Wang2015a, Karpiak2019, Ghiasi2020}, or into TDMCs \cite{Zollner2020, Zhong2020}, furthermore, a strong proximity effect on correlated magic angle twistronics devices \cite{Arora2020, Lin2102.06566}, or to stabilize fragile states like the topological insulator phase in graphene-TDMC heterostructures \cite{Island2019}.

%% file: sec_ack_contrib.tex
\section{Author contributions}

S.Z, M.K and P.M. fabricated the devices. 
Measurements were performed by B.F., A.M. with the help of E.T., B.Sz. 
B.F. did the data analysis.  
M.G. did the theoretical calculation. 
B.F. and P.M. and Cs.Sz. wrote the paper and all authors discussed the results and worked on the manuscript. 
K.W. and T.T. grew the hBN crystals. 
The project was guided by Sz.Cs., P.M, C.S. and J.F.

\section{Acknowledgments}

This work acknowledges support from the Topograph FlagERA network, the OTKA FK- 123894 grants, the Swiss Nanoscience Institute (SNI), the ERC project Top-Supra (787414), the Swiss National Science Foundation, the Swiss NCCR QSIT.  
This research was supported by the Ministry of Innovation and Technology and the National Research, Development and Innovation Office within the Quantum Information National Laboratory of Hungary and by the Quantum Technology National Excellence Program (Project Nr. 2017-1.2.1-NKP-2017-00001), by SuperTop QuantERA network, by the FET Open AndQC netwrok  and Nanocohybri COST network. 
P.M. and E.T. received funding from Bolyai Fellowship. 
M.G. acknowledges Scientific Grant Agency of the Ministry of Education of the Slovak Republic under the contract No. VEGA 1/0105/20.
K.W. and T.T. acknowledge support from the Elemental Strategy Initiative conducted by the MEXT, Japan, Grant Number JPMXP0112101001, JSPS KAKENHI Grant Numbers JP20H00354 and the CREST(JPMJCR15F3), JST. 

The authors thank Andor Kormányos, András Pályi and Péter Boross for fruitful discussions, and Márton Hajdú, Ference Fülöp team for their technical support.

%% file: references.tex
\bibliography{D:/in_this_folder}
\bibliographystyle{ieeetr}

%% file: SM_raw.tex
\section{Raw measurement data}

\begin{figure}[h]
\includegraphics[width=0.9\textwidth]{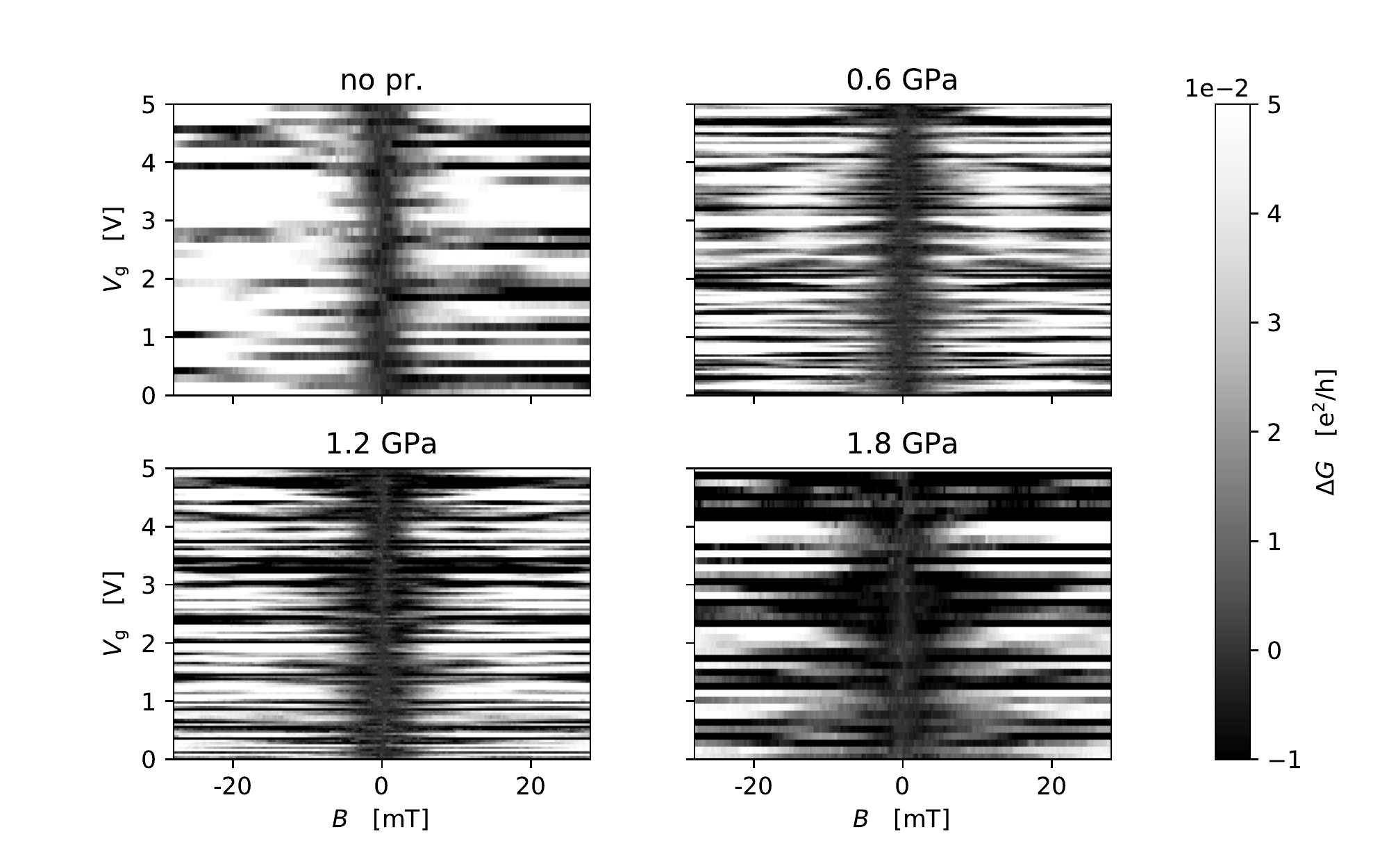}
\caption{
Conductance correction vs. gate voltage $V_\tn{g}$ and magnetic field $B$. 
The colorbar and the axes values are the same for each map.
A vertical gray line $\Delta G = 0$ at $B = 0$ is visible on all maps due to the correction method.
A darker gray shoulder becomes more and more visible along the $V_\tn{g}$ axis as pressure rises at low magnetic fields ($B<5\unit{mT}$).
At higher magnetic fields ($B > 5 \unit{mT}$) universal conductance fluctuations (UCF) become dominant leading to irregular magneto-conductance curves.
The colorbar is set to resolve the peak appearance in the low magnetic regime, thus the signal of UCF saturates it.
}
\label{fig:SM_raw}
\end{figure}

In this section, we present in Figure \ref{fig:SM_raw} all the symmetrized magneto-conductance measurement data used for the analysis without the averaging over the gate voltage range.
A vertical grey line is present at $B=0$ due to the correction $\Delta G(V_\tn{g}, B) = G(V_\tn{g},B) - G(V_\tn{g}, B=0)$.
As the pressure increases, a dark grey shoulder appears on both sides of this line corresponding to a growing WAL peak in the magneto-conductance.

%% file: SM_compare.tex
\section{Comparison of pressures at different gate voltages}

In Figure S2, we present detailed comparison of the magneto-conductance curves at different gate voltages.
Each curve is made by an averaging of all the recorded curves within a $\pm 1.5\unit{V}$ range around the value indicated in the panel title.
It is visible that the growing tendency of the WAL peak is present at all gate voltages and therefore this is a robust result of the pressure change.

\begin{figure}[h]
\includegraphics[width=\textwidth]{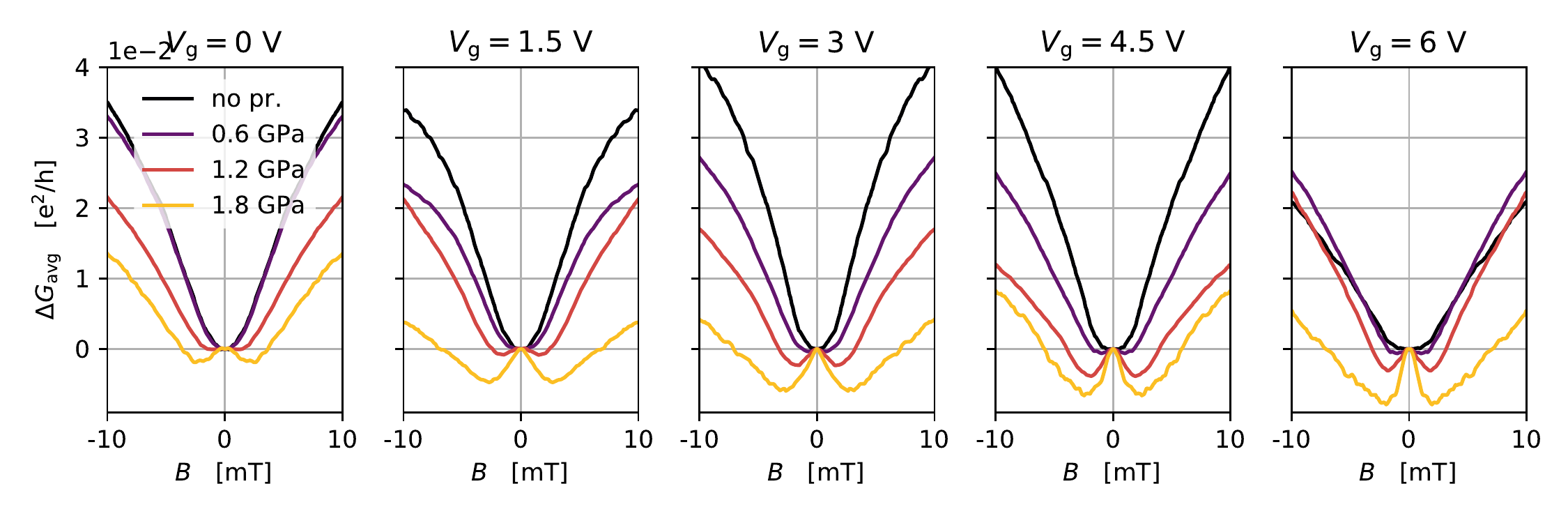}
\caption{
Comparison of magneto-conductance curves at different gate voltages.
Each curve is made by an averaging of all the recorded curves within a $\pm 1.5\unit{V}$ range around the value indicated in the panel title.
The colors mark the same pressure on each plot.
The WAL peak is more pronounced at all gate voltages as the pressure is increased, which suggests that the increasing pressure has an influence on the enhanced WAL feature independently on other, gate-dependent scattering mechanisms.
Conductance correction above $5\unit{mT}$ includes a pronounced contribution of UCF which leads to irregular curve shapes.
In addition, gate voltage dependence of the line shape is also clearly visible.
The peak amplitude grows with the gate voltage, which is in agreement with Ref.\,\cite{Tikhonenko2008} and can be explained with the change of other time scales along with the ones associated to SOC.
This tendency is present at every pressure.
Due to this dependence on gate voltage, one has to find a balance between reducing the effect of UCF sufficiently and not to average over a gate voltage range where the physical processes differ too much.
After investigating the gate voltage measurements and magneto-conductance curves averaged over different gate voltage ranges, we found that the curves of $4 \unit{V} \pm 1.5 \unit{V}$ give a good compromise.
This is the setting we used for the fitting.
}
\label{fig:SM_compare}
\end{figure}

%% file: SM_tdep.tex
\section{Temperature dependence}

Here, we present in Figure S3 the temperature dependence of the symmetrized magneto-conductance curves at maximum pressure with averaging between 0\,V and 2\,V gate voltage.
It is visible that the WAL peak disappears and the background is negligible above 30\,K at low magnetic fields.

\begin{figure}[h]
\includegraphics[width=10cm]{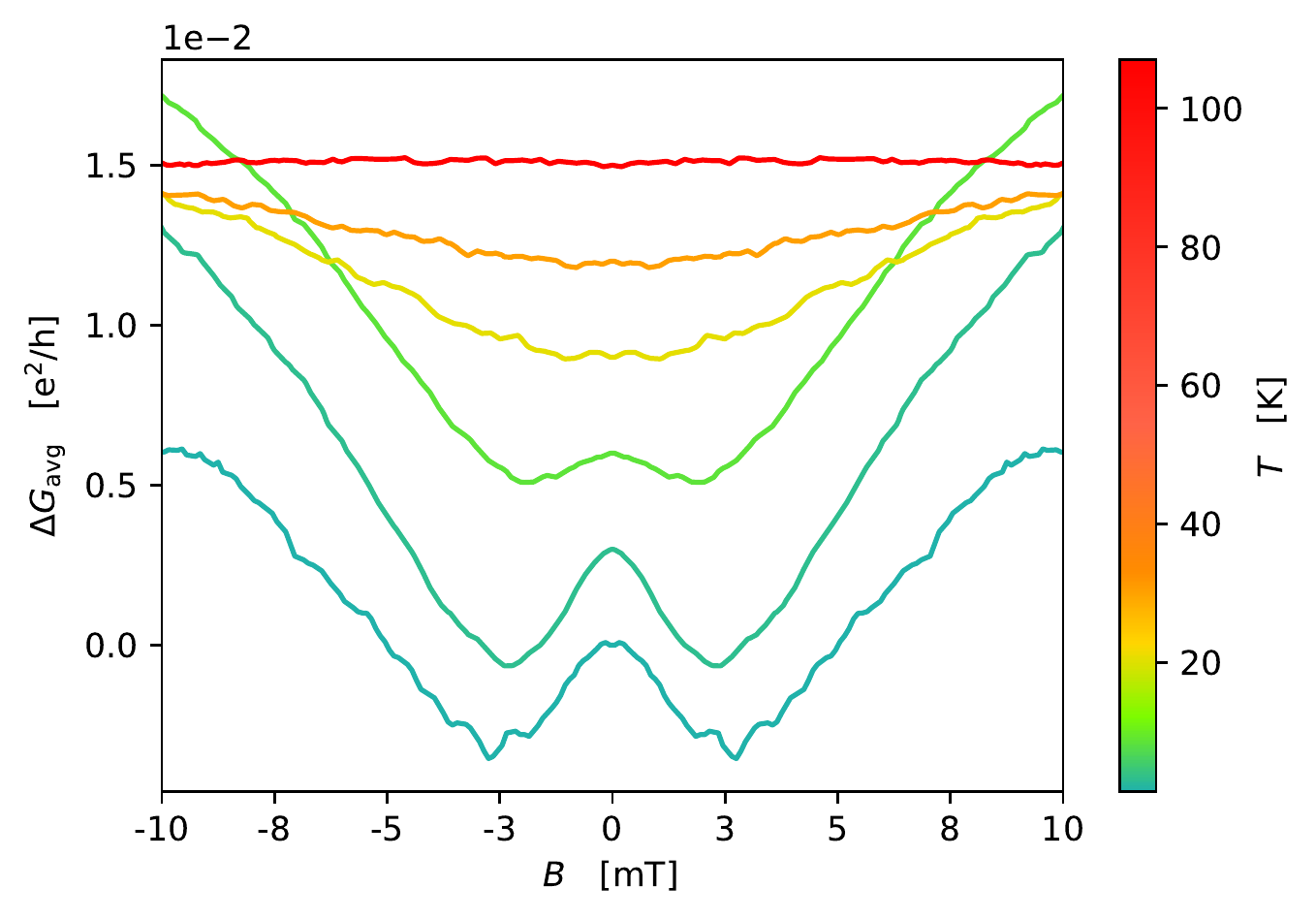}
\caption{
Temperature dependence of the WAL signal at $p = 1.8\unit{GPa}$.
The amplitude of the WAL peak decreases with increasing temperature due to the reduction of $\tph$.
The curves are shifted by $0.003 \, e^2/h$ for clarity.
}
\label{fig:SM_tdep}
\end{figure}

%% file: SM_fitting.tex
\section{Fitting method}

If the scattering times of the SOC ($\ta, \ts$) are comparable to the intervalley scattering time  ($\tiv$), both SOC and chirality influences the magneto-conductivity at the same B field scale, and a unified formula is needed to describe the weak localization behaviour.
This can be derived from Ref.\,\cite{McCann2012}, as has been done in Ref.\,\cite{Zihlmann2018}:

\begin{equation}
\label{eq:soc5}
\begin{split}
\Delta \sigma(B) = -\frac{1}{2} \frac{e^2}{\pi h} 
\left[ 
		F \left( \frac{\tbi}{\tphi} \right)	\right.
		& - F \left( \frac{\tbi}{\tphi + 2 \tai} \right)
		- 2 F \left( \frac{\tbi}{\tphi + \tai + \tsi} \right)	\\	
		& - F \left( \frac{\tbi}{\tphi + 2 \tivi} \right)
		- 2 F \left( \frac{\tbi}{\tphi + \tivi + \tiai} \right)		\\
		& + F \left( \frac{\tbi}{\tphi + 2 \tivi + 2 \tai} \right)
		+ 2 F \left( \frac{\tbi}{\tphi + \tivi + \tiai + 2 \tai} \right)\\
		& + 2 F \left( \frac{\tbi}{\tphi + 2 \tivi + \tai + \tsi} \right)
		\left. 
		+ 4 F \left( \frac{\tbi}{\tphi + \tivi + \tiai + \tai + \tsi} \right)
\right],
\end{split}
\end{equation}

\noindent{}where $\tiai$ is a scattering rate due to the intravalley scattering due to soft potentials and to the trigonal warping, together. 
This 5-parameter formula simplifies to the 3-parameter formula used in the main text if $ \tivi, \tiai \ll \tphi, \tai, \tsi$.
Remarkably, this complete formula is invariant under the pairwise interchange of rates $(\tai, \tsi)$ and $(\tivi, \tiai)$, i.e.\ chirality and SOC contribute to the magneto-conductivity in the same manner.
We note that we present this form of the expression instead of using the commonly used substitution of $\tsti = \tivi + \tiai$ to emphasize the symmetry of SOC and scattering parameters.
This symmetry makes the interpretation of the fit variables ambiguous unless the value of at least one of the parameters can be determined via an independent measurement.
We assume that the momentum relaxation time $\tm$, extracted from the gate voltage measurements for each pressure, corresponds to the intravalley scattering time $\tia$, which is shorter than any other time scale.
Therefore, during curve fitting, the measured value of $\tm$ is used as the value of the fixed parameter $\tia$ leaving only 4 parameters to optimize; and as a lower bound for all other time parameters for both the 3-parameter and the 5-parameter formulas.

Due to the complex shape of the fit formulas, one can expect multiple fit parameter sets that approximate a measured signal reasonably well.
To assess the uncertainty associated to the fitting parameters of this origin, we introduce a fitting method called ``grid fit'' that is described in the following.
During the curve fitting of a single magneto-conductivity curve, we do not search for the global best fit, but we look for all the sufficiently good fits in the parameter space.
To this end, we create a grid of starting parameters in the 3D parameter space of $(\tph, \ta, \ts)$ for the 3-parameter formula, or the 4D parameter space $(\tph, \ta, \ts, \tiv)$ of the 5-parameter formula (as $\tia$ is fixed), and we start a Levenberg-Marquardt fitting algorithm at each grid point.
We accepted the fit if the reduced chi-squared statistic of the fit result was reasonably close to the best value observed in the whole grid.
The start points' grid extends for each time parameter from the actual value of $\tm$ for each pressure to an upper limit of $1\cdot 10^{-11}\unit{s}$, which we consider as a reasonable upper limit for $\tph$ at the measurement temperature and therefore for any other time scale that has a visible influence on the magneto-conductivity curves.
Since we looked at parameter values extending over multiple order of magnitude, the distribution of the grid points for each parameter was log-uniform, i.e.\ the values of $\ln \tau_\tn{i}$ were equidistant for $i \in \{\varphi, \tn{asy}, \tn{sym}\}$ in the studied range.

All the accepted fits approximate the measurement data equally well for both formulas, with point-to-point differences of the curves 3 orders of magnitude smaller than the measurement values.
In Figure S4b-c, a summary of the fit parameters is shown as an example for the grid fit of the 0.6\,GPa curve using the 5-parameter formula.
The result of each executed fit is represented by a pair of points on the two panels.
The position of the points correspond to the fit values of the parameters marked on the axes, and the red color marks the accepted fits.
Fits that were a result of the Levenberg-Marquardt algorithm started from a parameter grid point but got rejected due to their high reduced chi-squared statistic are marked by blue.
For the parameters $\tph, \ta, \tiv$ the accepted fit values are concentrated on a small part of the studied range which suggests reliable fit values. 
However, similarly to the results using the 3-parameter formula, the value of $\ts$ can vary across orders of magnitude without decreasing the fit goodness, meaning that the valley--Zeeman effect does not have any measurable effect on the weak localization behavior in this experiment.
Therefore we conclude that the method can not extract reliable value for $\ts$.

\begin{figure}
\includegraphics[width=\textwidth]{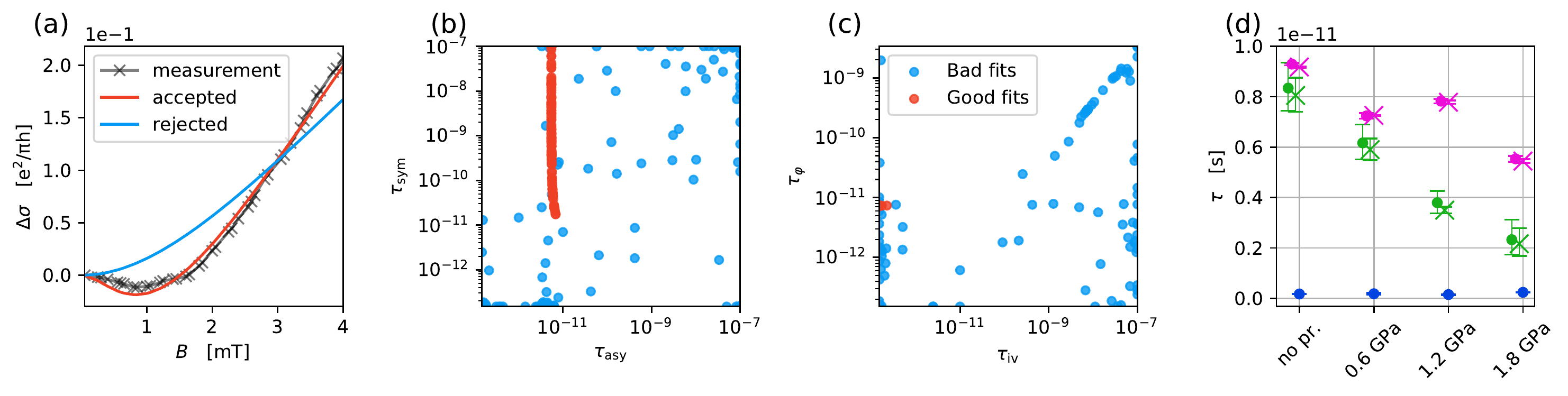}
\caption{
(a)
Example of accepted (red) and rejected (blue) fit results using the 3-parameter formula on the 0.6\,GPa curve.
Fit results were accepted based on their reduced chi-squared statistic.
(b-c)
Summary plots of the fit parameters of the accepted fits to the 0.6\,GPa measurement data curve using the complete formula, shown as an example.
Every result of the Levenberg-Marquardt method is represented by a pair of dots at the fit value of all the 4 fit parameters.
Red dots represent fits that approximate the measurement data reasonably well.
Blue dots represent the rejected fit results.
(d)
Comparison of fit parameters using the formulas.
The pink color corresponds to $\tph$, the green one to $\ta$.
The cross marker is the fit value using the 3-parameter formula, the dot using the 5-parameter formula. 
The errorbars represent the minimum and maximum values, the data point is the geometric mean value of the accepted fits.
Using the 5-parameter formula, both $\tph$ and $\ta$ obtain values very close to the results of the 3 parameters formula.
The values for $\tiv$ using the 5-parameter formula are also plotted with blue.
Its small values compared to $\tph$ and $\ta$ validates the applicability of the 3-parameter formula.
The data points are slightly shifted horizontally to avoid overlapping errorbars.
}
\label{fig:curvefit}
\end{figure}

\begin{figure}
\includegraphics[width=6.5cm]{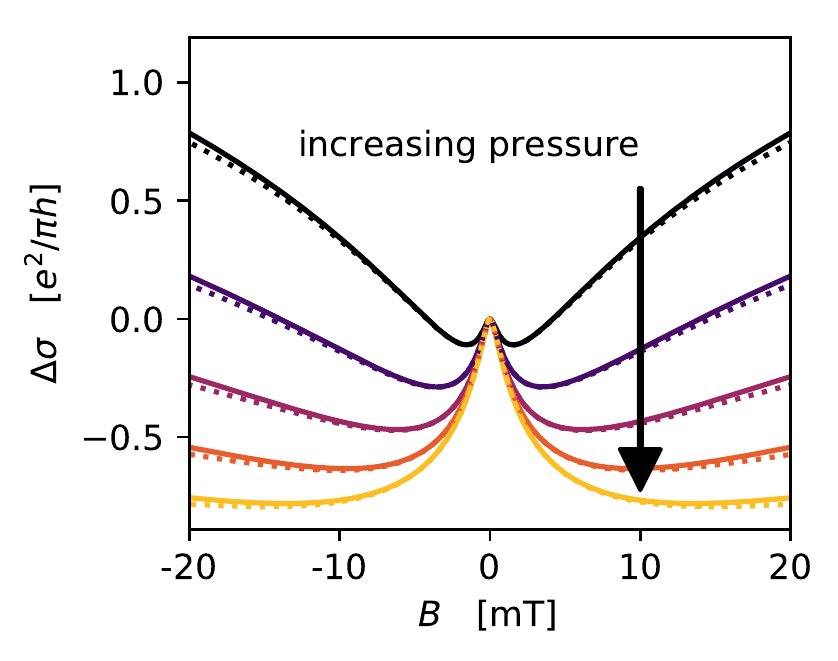}
\caption{
Simulated weak anti-localization curves using both the 3-parameter (solid line) and 5-parameter (dotted line) formulas. 
The curves presented in the main text correspond to the solid lines.
The difference of the two curve set is only visible at high magnetic fields, the shape of the central peak is unaffected.
Therefore, the two formulas assess the spin orbit times similarly, which supports the parameter fit values in Figure S4d.
}
\label{fig:simul}
\end{figure}

For the other parameters, we consider all the accepted outcomes as physically valid values, since these parameter sets are equivalent in the curve shape.
We attribute the geometric mean of the maximum and the minimum of the accepted values as the fit value of the parameter and the upper and lower error boundaries are the minimum and maximum of the accepted outcomes.
The results are summarized in Figure S4d.
The fit values of $\tph$ and $\ta$ are very close for both formulas and show the decreasing trend discussed in the main text.
The values of $\tiv$ are indeed much smaller than the other time scales, which suggests that the influence of chirality is indeed negligible in our case and the 3-parameter formula is applicable.
This is also demonstrated in Figure S5, where 5 pair of simulated curves of the same parameters are plotted using the 3-parameter and the 5-parameter formula.
The 3-parameter curves are also shown in Figure 1b of the main text.
The $\ta$ and $\ts$ parameters were calculated from the $\LR$ and $\LVZ$ parameters, respectively.
For the simulation of the pressure, $\LR$ was changed from 300\,\textmu{}eV to 900\,\textmu{}eV in 5 equidistant steps, whereas $\LVZ$ was changed from 800\,\textmu{}eV to 2.0\,meV in the same manner.
The other parameters were fixed for all curves both for the calculation of the magneto-conductivity curves and of the SOC time parameters: $\tph = 1\cdot10^{-11} \unit{s}$, $\tiv = 1 \cdot 10^{-13} \unit{s}$, and $\tia = 1 \cdot 10^{-13} \unit{s}$.
The difference between the 3-parameter and the 5-parameter curves is only visible at higher magnetic fields.

%% file: SM_otherdevice.tex
\section{Measurements on another device}

Here, we present measurement results on device B.
This heterostructure consists of the same layers as device A, and the measured two-terminal segment was $L = 5$\,\textmu{}m long and $W = 1$\,\textmu{}m wide.
For this device, a growing WAL peak is also present in the magneto-conductance curves with increasing pressure.
However, due to gate instabilities and drifts, a proper comparison of the magneto-conductance curves was not possible.

\begin{figure}[h]
\includegraphics[width=10cm]{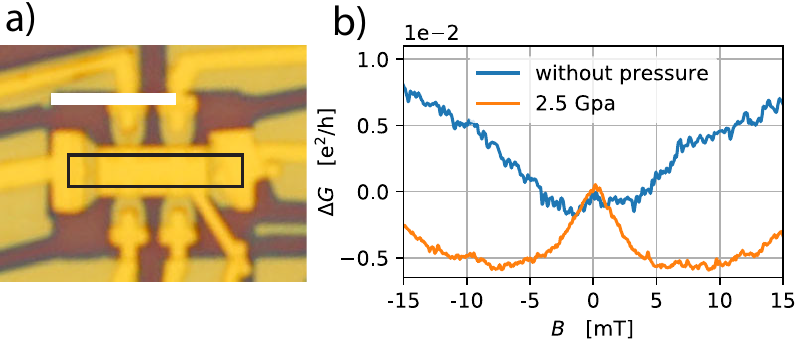}
\caption{
Measurements of device B.
(a)
Optical micrograph of the device with the studied segment highlighted by black rectangle.
$L = 5$\,\textmu{}m, $W = 1$\,\textmu{}m.
4 terminal voltage was measured on the electrodes of the lower side of the picture, but not used during the analysis discussed here.
The electrodes on the upper side were left floating.
(b)
Two terminal magneto-conductance signal, averaged between +2\,V and +4\,V from the charge neutrality point.
This device also exhibits an increasing WAL peak at higher pressures, but a proper comparison of the pressures was not possible due to gate instabilities.
}
\label{fig:SM_otherdevice}
\end{figure}